# PERFORMANCE EVALUATION OF REALISTIC VANET USING TRAFFIC LIGHT SCENARIO

Nidhi[1] and D.K. Lobiyal[2]

[1,2]School of Computer and Systems Sciences, Jawaharlal Nehru University, New Delhi, India
`nidhi.jjmsn@gmail.com`, `lobiyal@gmail.com`

## ABSTRACT

*Vehicular Ad-hoc Networks (VANETs) is attracting considerable attention from the research community and the automotive industry to improve the services of Intelligent Transportation System (ITS). As today's transportation system faces serious challenges in terms of road safety, efficiency, and environmental friendliness, the idea of so called "ITS" has emerged. Due to the expensive cost of deployment and complexity of implementing such a system in real world, research in VANET relies on simulation. This paper attempts to evaluate the performance of VANET in a realistic environment. The paper contributes by generating a real world road Map of JNU using existing Google Earth and GIS tools. Traffic data from a limited region of road Map is collected to capture the realistic mobility. In this work, the entire region has been divided into various smaller routes. The realistic mobility model used here considers the driver's route choice at the run time. It also studies the clustering effect caused by traffic lights used at the intersection to regulate traffic movement at different directions. Finally, the performance of the VANET is evaluated in terms of average delivery ratio, packet loss, and router drop as statistical measures for driver route choice with traffic light scenario. This experiment has provided insight into the performance of vehicular traffic communication for a small realistic scenario.*

## KEYWORDS

*Intelligent Transportation System, Vehicular Ad-hoc Networks, Geographical Information System, Mobility Model Generator for Vehicular Networks[v.2.9], Simulation of Urban Mobility[0.12.3], Network Simulator-2.34.*

## 1. INTRODUCTION

As per the World Health Organization (WHO) statistics, more than 1.3 million people worldwide are estimated to be killed each year out of road accidents. According to an online article published in Deutsche Welle [2] by Murali Krishnan dated 29.04.2010, "India's record in deaths has touched a new low, as toll rose to at least 14 deaths per hour in 2009 against 13 the previous year". While trucks/lorries and two-wheelers were responsible for over 40% deaths, the rush during afternoon and evening hours were the most fatal phases.[2,3]. Also, as per another article of WHO (article in Times of India, Dipak Kumar Dash, TNN, Aug 17, 2009, 04.10am IST) India leads the world in road deaths. In addition to this, some of the common problems to tackle with are the "Miles of Traffic Jam" on highway and the "Search for best Parking Lot" in an unknown city. [1]

For all the above mentioned reasons, the Government and Automotive Industries today pay lot of attention towards traffic management and regulation of a smooth traffic. They are investing many resources to slow down the adverse effect of transportation on environment, thereby increasing traffic efficiency and road safety. The advancements in technology, in the areas of Information and Communications, have opened a new range of possibilities. One of the most promising areas is the study of the communication among vehicles and Road Side Units (RSUs), which lead to the emergence of Vehicular Network or Vehicular Ad-hoc Network (VANET) into picture. [1,4].

VANET is characterized as a special class of Mobile Ad hoc Networks (MANETs) which consists of number of vehicles with the capability of communicating with each other without a fixed infrastructure. The goal of VANET research is to develop a vehicular communication system to enable 'quick' and 'cost-efficient' transmission of data for the benefit of passenger's safety and comfort. Due to the expensive cost of deploying and complexity of implementing such a system in real world, research in VANET relies on simulation. However, the simulation depends on the mobility model that represents the movement pattern of mobile users including its location, velocity and acceleration over time. A mobility model needs to be a Realistic Mobility Model that considers the characteristics of the real world scenario either by taking a real world MAP obtained from TIGER(Topologically Integrated Geographic Encoding and Referencing) database from U.S. Census Bureau or by taking Satellite images of Google Earth into consideration to simulate a realistic network.[1]

In V2V communication or Inter-Vehicle communication, Vehicles are able to communicate with other ongoing vehicles on their path . In this scenario, it is not known in advance when it is possible to meet another vehicle to which the communication is feasible. In V2I communication, vehicles are able to communicate with Road Side Unit (RSUs) or access points. Whereas, In Inter-Road Side Communication it is possible to know the communicating parties in advance as RSUs are placed at a fixed distance from each other. Therefore, the main difference between V2V and V2I is the coverage area[8,21].

Rest of the paper is organized as follows. Related work is briefly described in Section 2. In Section 3, the methodology of proposed work is explained along with various tools which are used to carry out the work. Section 4 further discusses the simulation of network, results and analysis obtained through simulations conducted. Finally, Section 5 concludes the work presented in this paper.

## 2. RELATED WORK

Research is being carried out in the field of VANET such as Analyzing data dissemination in VANETs, Identifying and studying routing protocols in VANET in terms of highest delivery ratio and lowest end-to-end delay etc. The issues of Security and Privacy also demands great attention. The study of Mobility Models and their realistic vehicular model deployment is a challenging task.[8] Random way Point(RWP)[9] is an earlier mobility model widely used in MANET in which nodes move freely in a predefined area but without considering any obstacle in that area. However, in a VANET environment vehicles are typically restricted by streets, traffic light and obstacles. Ana et. al. [20] considers the mobility model in which vehicles know from the start their initial and final points. The routing track is then chosen considering the social relation between the vehicles and also the destination point. This means that vehicles move only between the initial and final point on path chosen by the social relation strength between the vehicles. GrooveSim [10] was the first tool for forecasting vehicular traffic flow and evaluating vehicular performance. It gives a traffic simulator environment which is easy to use for generating real traffic scenario for evaluation. But it fails to include network simulator as it was unable to create traces for network. David R. Choffnes et al. [11] proposed a mobility model named STRAW (Street RAndom Waypoint). This model has taken real map data of US cities and considered the node (vehicle) movement on streets based on this map. This model also has the functionality to simplify the traffic congestion by controlling the vehicular mobility. But still it lacks overtaking criteria that cause convey effect in street as it considered random method which is not realistic. Kun-chan Lan et al.[6] describes a realistic tool MOVE for generating realistic vehicular mobility model. It is built on top of an open source micro-traffic simulator SUMO and its output is a realistic mobility model that can immediately be used by popular network simulators such as ns-2 and qualnet. In paper [1] Kun-chan's MOVE model (v.2.81) was used to obtain the results of driver's route choice at the intersection without considering the traffic light scenario.

## 3. PROPOSED WORK & METHODOLOGY

To evaluate the performance of VANET, there is a need to deploy a real world scenario with all the vehicular constraints. In this paper the experiment was performed by taking a limited bounded region of a real world scenario i.e. "JAWAHARLAL NEHRU UNIVERSITY (JNU), NEW DELHI, INDIA" into consideration. The steps to implement a VANET simulation in this region are as follows:

- Generation of JNU Map
- Creation of Vehicular Traffic flow on this Map
- Simulation of Network with traffic lights at Intersections

The detailed procedure in implementing such above mentioned steps are explained in the rest of this paper.

### 3.1. JNU Map Generation

For creating a real world Map of JNU, Some of the existing tools have been used such as Google Earth, ArcGIS 9 (ArcMap version 9.1), MOVE Simulator (v 2.9)[5,6] and Adobe Dreamweaver CS4. Satellite image of JNU has been taken from Google Earth shown in Figure 1. This image was further imported into ArcGIS 9 as depicted in Figure 2. **ArcGIS** is basically a suite consisting of a group of Geographic Information System (GIS) software products.[12]

**NOTE:** Google Earth gives latitude and longitude of a particular location whereas ArcGIS maps those latitudes and longitudes to the required coordinate plane with the desired origin in a Two Dimensional Space.

Some of the 2-D Co-ordinates of this Map were not lying in the first quadrant of the 2-D Co-ordinate plane. In order to obtain all the co-ordinates in the first quadrant, the origin was shifted to an appropriate location. Shifting of the old Co-ordinates (x, y) to a new origin (h, k) is given by :

$$X = x + h; Y = y + k ;$$

Where (X,Y) represents the translated Co-ordinates in the plane with new origin and the traffic lights at particular co-ordinates of intersection have been used as the inputs to the Map Node Editor of MOVE Simulator as shown in Figure 3. After creating nodes and traffic lights at particular intersections using Map Node editor, numbers of parameters are defined such as edges between nodes, number of lanes, speed and priority of roads on which vehicle move, with the help of Road Editor of MOVE simulator as shown in Figure 4. Here, a multi-lane scenario of two lanes with 75% road priority has been set. The threshold speed has been considered for each lane in a region of JNU Map as 40m/s. Next a connection was established between nodes via edges by writing an XML code (**.con.xml**)[16] using Dreamweaver CS4. Finally the nodes, edges and connection files are configured into **.net.xml** by using NETCONVERT to create the MAP. Figure 5 depicts the JNU Map created by the above defined tools.

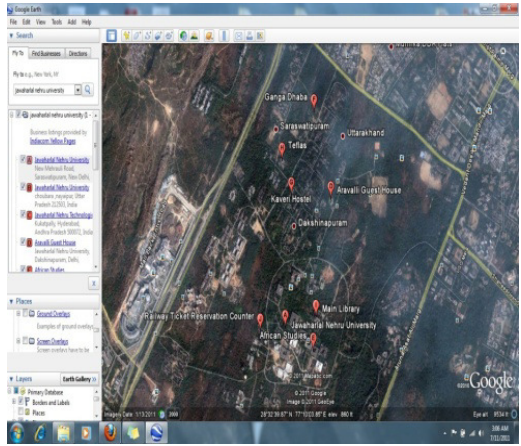
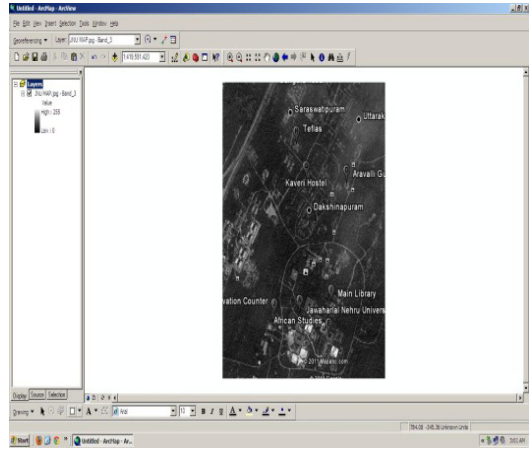

Figure 1 Satellite Image of JNU.    Figure 2 Imported Image of JNU in ArcGIS

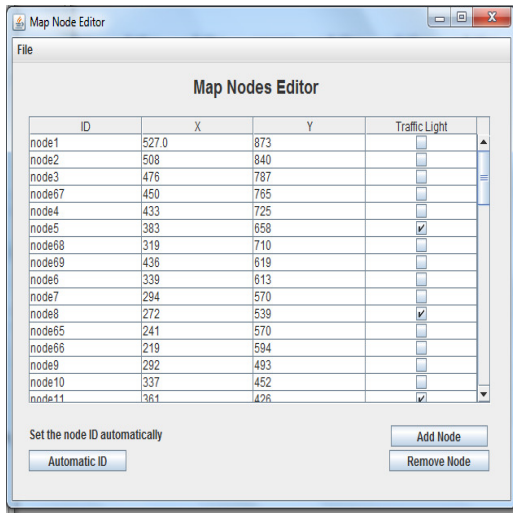
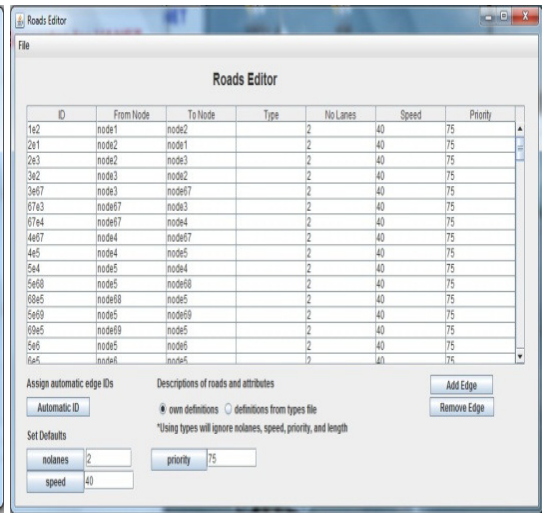

Figure 3. Map Node Editor of MOVE    Figure 4. Road Editor of MOVE

### 3.2 Traffic Flow

For generating a vehicular traffic flow by taking traffic lights into consideration on the above created Map, SUMO 0.12.3[17] simulator has been used in addition to MOVE simulator. Initially, the Route File in XML (**rou.xml**) was created, in which acceleration, deceleration, maximum speed, length and type of a vehicle were specified (see Table 1). In addition to this, the bounded JNU region has been divided into 36 smaller routes which the vehicles can take. Further, the departure time of a particular vehicle on a particular route which creates the vehicular traffic flow among the nodes has been specified. The vehicle's destination from the source and their turning directions at the intersections, such as right turn, left turn and straight as per their destination were also set as per the driver's route choice at intersection. In addition to this, traffic light scenario has been taken into consideration at all the intersections to regulate the traffic movement in different direction as well as to analyze the clustering effect at the intersections.

Table 1 Types of Vehicle and their Characteristics

| Vehicle Type | Max. Acc. (m/s$^2$) | Max. Dec. (m/s$^2$) | Length (m) | Max. Speed (m/s) |
|---|---|---|---|---|
| Car A | 3.0 | 6.0 | 5.0 | 30 |
| Car B | 2.0 | 6.0 | 7.5 | 30 |
| Car C | 1.0 | 5.0 | 5.0 | 20 |
| Car D | 1.0 | 5.0 | 7.5 | 10 |

Different Route files with traffic lights at the intersections have been created for varying traffic flow consisting of 10,20,30,40,50,60,70 vehicles. This varying flow has been set by keeping in mind, a constant deceleration and acceleration model in which vehicles do not move and stop abruptly. Map file (.net.xml) and the different Route files (rou.xml) of varying traffic flow were configured to create the corresponding trace files (**sumo.tr**) which can be visualized using SUMO simulator. These trace files basically shows the JNU Map as shown in Figure 5 and the flow of traffic and clustering effects of vehicles due to traffic lights at the intersection are depicted in Figure 6(a) and (b). After setting the parameters of SUMO, the real world scenario of JNU region can be visualized with vehicles moving on it as depicted in Figure 6 (a), (b) & (c).

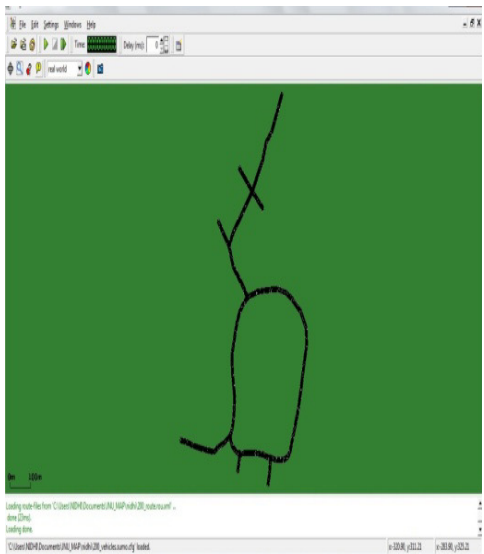
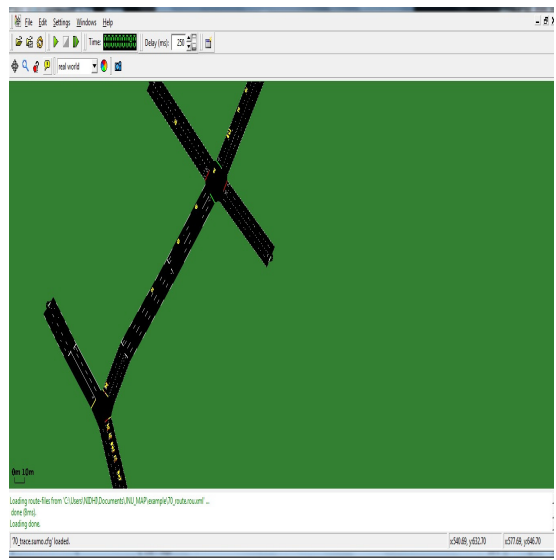

Figure 5. SUMO visualization of JNU Map .     Figure 6(a). Vehicular Flow and Clustering of vehicles due to traffic light at the intersection

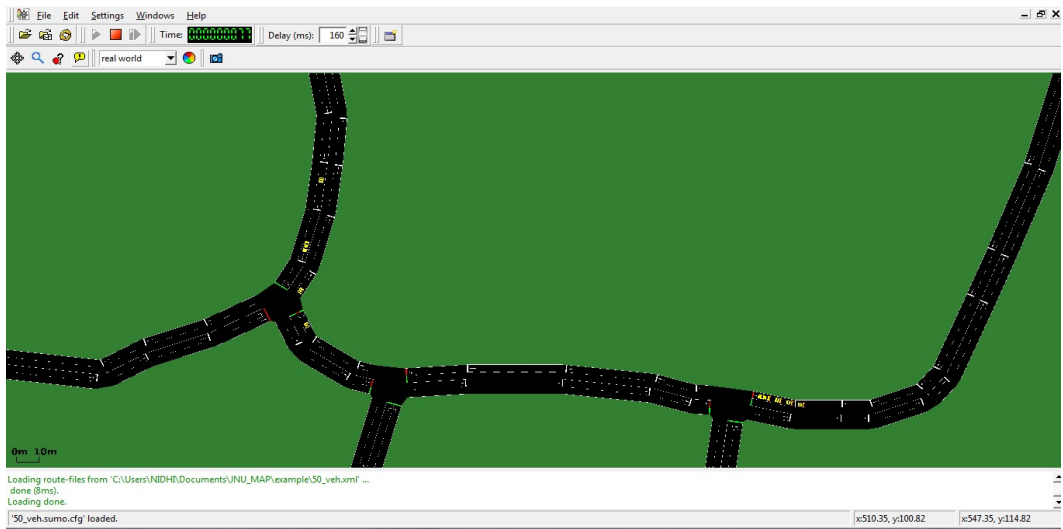

Figure 6(b)

## 4. TRAFFIC LIGHT SIMULATION

In order to simulate the clustering effects of vehicles, traffic lights have been installed in the network at the intersections shown in Figure 6(a) and (b). The traffic flow among the vehicles is generated using Traffic Model Generator of MOVE and Network Simulator (NS-2.34)[18].

The Traffic Model Generator of MOVE creates the dynamic mobility of varying number of vehicular traffic by generating traffic simulation file for simulation. The traffic simulation files have been generated by interfacing traffic flow with traffic lights created in section 3.2 with the JNU Map created in section 3.1. These traffic simulation files of different traffic scenario are subsequently used in NS-2 which facilitated the simulation of traffic flow of region under study.

Various parameters were considered for establishing the communication among vehicles. For example, a vehicular traffic flow was deployed using IEEE 802.11 standard with transmission range of 250 meters. The other parameters used are discussed in Table 2.

**Table 2.** Network parameters

| Parameters | Values |
| --- | --- |
| Channel Type | Wireless Channel |
| Propagation Model | Two Ray Ground Model |
| Network Interface Type | Wireless Phy |
| MAC Type | 802.11 |
| Interface queue | DropTail/Pri Queue |
| Link Layer Type | LL |
| Anetnna | Omni Antenna |
| Ifqlen | 50 |
| Varying No. of Nodes | 10,20,30,40,50,60,70 |

| Routing Protocol | AODV |
| --- | --- |
| Topology (X,Y) Co-ordinates | (659, 911) |
| Transmit Power, Pt | 0.2818 |
| Channel Frequency | 2412e+6 |
| RXThresh | 3.65262e-10 |
| CSThresh | (Expr 0.9 * RXThresh) |

The simulation covers 600349 m$^2$ area and the following parameters have been used for traffic flow between nodes in our simulations.
.

Table 3. Parameters of Traffic Flow between nodes

| **Parameters** | **Values** |
| --- | --- |
| Agent | UDP |
| Packet_size | 1000 |
| Application_Traffic | CBR |
| CBR Rate | 64kbps |
| CBR_max_pkts | 2280000 |
| CBR interval | 0.05micro sec |
| Different RNG seed | 2,4,6,8,10 |

After setting up the network and traffic flow as discussed above, simulation was conducted by taking 1/4$^{th}$ of the vehicular nodes sending CBR traffic for a traffic scenario of 10 vehicles initially. In this scenario, 50% of the vehicles are involved in direct communication, whereas, rest 50% vehicles serves as the intermediate (or router) nodes for the communication. Further, all the traffic parameters as given in table 3 were kept constant except number of nodes sending CBR traffic. This we have considered always as 1/4$^{th}$ of the number of vehicles 20,30,40,50, 60 and 70 respectively.

### 4.1 Simulation Results

The impact of realistic vehicular mobility (using various tools as discussed in Section 3), on the performance of ad-hoc routing protocols has been evaluated in this section.

The driver route choice behaviour with traffic lights at the intersections has been simulated for a real world scenario. In this, all possible routes from the source to destination are defined and the driver needs to decide about which route is to be taken from among all possible routes at any intersection. The presence of traffic lights at the intersection regulates the smooth movement of vehicles in different directions and causes clustering effect by forcing the vehicles to stop at intersection when the signal is red. Therefore, the node density at the intersection increased which improves the network connectivity among the peers at intersection, but the improved connectivity deteriorates the packet delivery ratio.

Our simulation concentrates on selecting the probability of choosing a route at the intersection. This probability directly determines the number of vehicles on a particular route. The data in terms of packets are transmitted to facilitate communication among vehicles. In order to study the behaviour of communication, the parameters like delivery ratio, packet loss and router drop has been considered which are discussed in the subsequent sections.

### 4.1.1 Average Delivery Ratio

Delivery Ratio implies the ratio of number of packets successfully delivered to the number of packets sent.
For calculating delivery ratio with respect to the number of vehicles, different traffic scenarios were simulated with varying number of vehicles in multiples of 10. For each scenario, delivery ratio was calculated for 5 simulation runs by changing the seed in multiples of 2. The Average delivery ratio for each scenario was an average
of 5 simulation runs and it is calculated as follows:

$$APR = (\sum_{k=1}^{5} PR) / 5$$

$$APS = (\sum_{k=1}^{5} PS) / 5$$

$$ADR \% = (APR/APS) * 100$$

Where, PR = Packet Received, PS = Packet Sent, APR = Average Packet Received and APS = Average Packet Sent.

The summary of results obtained is shown in table 4 and the results are further analyzed graphically in figure 7. It can be observed that the choice of route at intersection points can significantly affect the simulation results.

Table 4 Number of Traffic and Avg. Delivery Ratio (ADR) %

| No. of Vehicular Traffic | Packet Delivery Ratio % |
|---|---|
| 10 | 87.68 % |
| 20 | 82.05 % |
| 30 | 76.76 % |
| 40 | 57.13 % |
| 50 | 52.42 % |
| 60 | 46.02 % |
| 70 | 16.61 % |

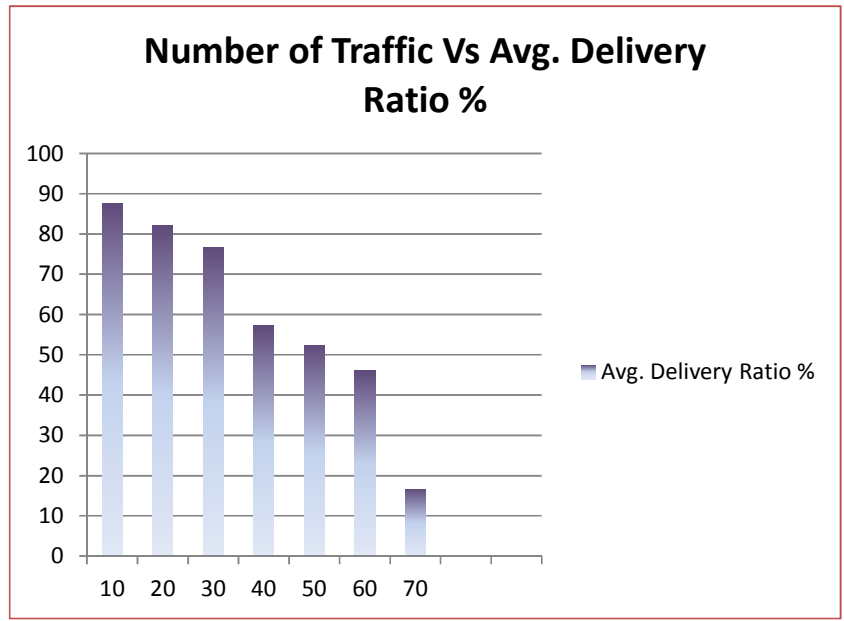

Figure 8. Number of Traffic Vs Avg. Delivery Ratio% using traffic light scenario.

It is clear from the figure 8 that the delivery ratio decreases as the node density increases due to the the increased collision caused by higher node density. However, as higher node density increases connectivity, but at the same time, it may result in more number of collisions.

### 4.1.2 Router Drop:

Router Drop for each traffic scenario is calculated by taking the average of Router Drop to the packets sent as shown below :

$$RD\% = \left(\sum_{K=1}^{5} \frac{RD}{PS}\right) * 100$$

Where, RD % = Router Drop %

### 4.1.3 Packet Loss:

Packet loss is calculated by taking the average of packet loss to the packets sent.

$$PL = (PS - PR)$$

$$PL\% = \left(\sum_{K=1}^{5} \frac{PL}{PS}\right) * 100$$

Where, PL = Packet Loss.

The results obtained for Router Drop and Packet Loss are summarized in table 5 for varying vehicular traffic. This is further illustrated in figure9.

Table 5 (Number of vehicular traffic) Vs (Router Drop and Packet Loss%)

| No. of Vehicular Traffic | Router Drop % | Packet Loss % |
|---|---|---|
| 10 | 12.31 % | 12.32 % |
| 20 | 06.71 % | 17.95 % |
| 30 | 23.09 % | 23.23 % |
| 40 | 36.97 % | 42.86 % |
| 50 | 40.62 % | 47.58 % |
| 60 | 41.28 % | 53.98 % |
| 70 | 55.45 % | 83.39 % |

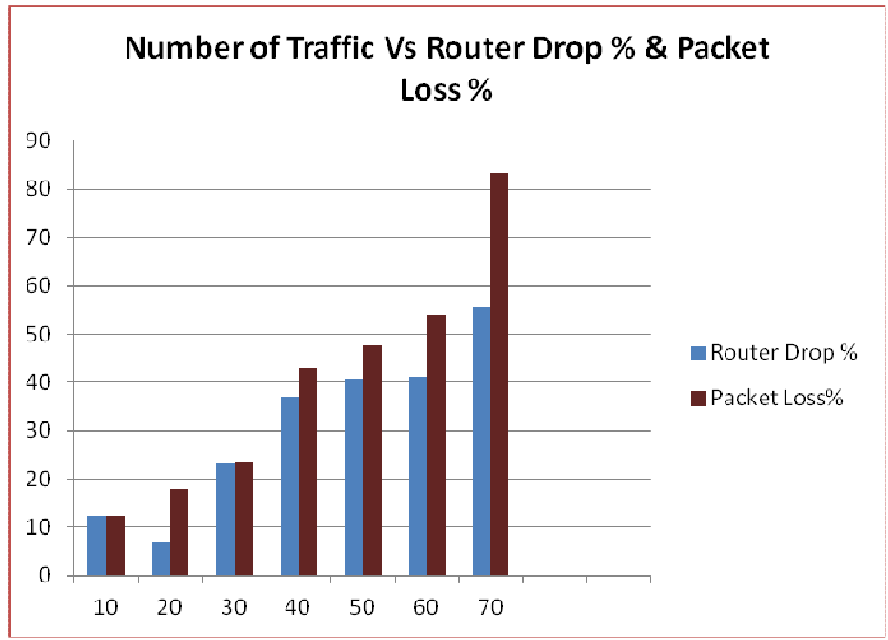

Figure 9. Vehicular Traffic Vs (Router Drop% & Packet Loss%) using traffic light scenario.

Our Simulation results suggest that, the presence of traffic lights at the intersections, deteriorate the packet transmission rate and increases the number of packet collision as shown in Figure 8 and 9. This happens due to the high node density at the intersection till the traffic light signal turns to green and it also results in collision of packets sent by vehicles at the same time. Further, it was observed that the packet delivery ratio kept on decreasing since the collision of packets completely depends on the clustering of vehicles at the intersection. This phenomenon can be explained by deployment and movement of vehicles in a given scenario. It seems that the connectivity between the vehicles get improved but simultaneously reduces the delivery ratio.

Figure 9 shows the effect of varying number of vehicles on packet loss and router drop together. A packet may be dropped by a vehicle or by a router. The packet loss percentage is considered as packets dropped by vehicles. Further in figure 9, it is quite evident that the percentage of packet loss was slightly more than router drops. This happens because of higher chances of packet being dropped at the end rather than being dropped at intermediate nodes. Here, again from the figure, it is quite evident that the router drop is not confined to any increasing or decreasing order. As explained for the case of lower deliver ratio in Figure 8, the drop rate was higher due to clustering effects of vehicles caused by traffic lights at the intersection.

## 5. CONCLUSION

In this paper, we have obtained an in-sight idea of simulating real world scenario of VANET. As it is not easy to deploy and implement such a complicated system in real world before knowing the impact of all parameters used in VANET, a small real world area i.e. our University, JNU itself, was taken into consideration, for studying the impact of mobility in the VANET. Traffic movement has been deployed across the area under consideration using one of the realistic vehicular mobility models. The behavior of this network was simulated using NS2 to study the impact of driver's choice with traffic lights at the intersection on packet transmission over V2V communication using AODV routing protocol and IEEE 802.11 standard.

The performance of the network has been evaluated by taking delivery ratio, packet loss and router drop as statistical measures. The average delivery ratio for various scenarios such as varying number of vehicles with constant power transmission range of 250m and frequency of 2.4GHz was observed to be 68.38% whereas packet loss was 40.18%.

Traffic light scenario has been an important measure to regulate the traffic flow in a round robin fashion. But for data transmission, it has become an obstacle since the packet forwarding nodes at the intersection drops the packets, due to the high number of transmission at the same time. Therefore, it is concluded from the results that packet drops may be reduced by using RSU's that can forward the packets even if the intermediate vehicles drops the packets at the intersections. Therefore, simulation of vehicular networks using RSU at the intersection will be our future direction of research.

**Authors**

*Nidhi* is pursuing Ph.D in Computer Science and Technology from School of Computer and Systems Sciences, Jawaharlal Nehru University (JNU), New Delhi, India. She pursued her M.Tech (Computer Science and Technology) from JNU, in 2011, B.Tech (Information Technology) from Guru Gobind Singh Indraprastha University, New Delhi, India in 2008 and Diploma in Computer Engineering from Jamia Millia Islamia University, New Delhi, India in 2005. Her area of interest includes Computer Networks, Ad-hoc Networks, and Vehicular Ad-hoc Networks.

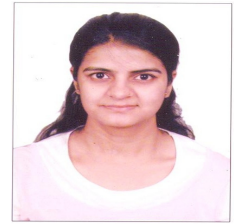

*Daya K. Lobiyal* received his Ph.D. and M.Tech. (Computer science) from School of Computer and Systems Sciences, Jawaharlal Nehru University, New Delhi, India in 1996 and 1991, respectively, and B.Tech. (Computer Science and Engineering) from Lucknow University, India in 1988. Presently, he is an Associate Professor in the School of Computer and Systems Sciences, Jawaharlal Nehru University, New Delhi, India. His research interest includes Mobile Ad hoc Networks, Vehicular Ad Hoc Networks, Wireless Sensor Network, Video on Demand and Natural Language Processing.

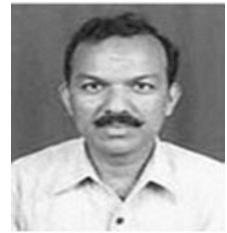